\documentstyle[pre,twocolumn,epsf,floats,aps]{revtex}

\newcommand{\beq}{\begin{equation}}    
\newcommand{\eeq}{\end{equation}}    

\begin{document}
\draft
 
\twocolumn[\hsize\textwidth\columnwidth\hsize\csname @twocolumnfalse\endcsname

\title{\LARGE Transport on an annealed disordered lattice}
\author{Radim Vo\v{c}ka}
\address{CEA, Service de Physique de l'Etat Condens\'e, 
F-91191 Gif-sur-Yvette, France 
}
\date{\today}
\maketitle
                                                                            
\begin{abstract}
We study the diffusion on an annealed disordered lattice with
a local dynamical reorganization of bonds. We show that 
the typical rearrangement time
depends on the renewal rate like $t_r \sim \tau^{\alpha}$ with $\alpha \neq 1$.
This implies that the crossover time to normal diffusion in a slow 
rearrangement regime shows a critical behavior at the percolation threshold.
New scaling relations for the dependence of the diffusion
coefficient on the renewal rate are obtained. 
The derivation of scaling exponents confirms
the crucial role of singly connected bonds in transport properties.
These results are checked by 
numerical simulations in two and three dimensions.
\end{abstract}

\vskip2mm ]

``Dynamic percolation''  (``stirred percolation'') 
\cite{Kutner:1983,Harrison:1985}
 was introduced as a model of transport in environments that
evolve in time, e.g. microemulsions or  
polymers (for further applications see \cite{Nitzan:1994}). The simplest
 version of the model is defined on a $d$-dimensional regular lattice. Each
pair of nearest neighbor sites  is connected by a bond, which can be
either conducting  or insulating. We note $p$ the proportion of conducting
bonds. Time evolution of the environment is achieved by a reorganization of
bonds, defined below. Diffusion of a tracer
particle in such a network is conveniently described by the 
ant-in-the-labyrinth paradigm \cite{deGennes:1976}. Two basic algorithms
are available. The ``blind'' ant chooses its direction randomly at each
time step  and moves only if 
the corresponding bond is conducting. The ``Myopic'' ant
chooses  among the  conducting bonds. Both algorithms lead
to the same scaling behavior of the diffusion coefficient. 
Two qualitatively different dynamic percolation models appeared in the 
literature. 
The global 
reorganization model is the simplest.  After some renewal time $T_r$,
the assignment of conducting bonds is updated throughout the lattice. The 
behavior of this model is well understood \cite{Nitzan:1994}, as it is 
closely related to the ordinary percolation. If  $\left<r^2\right>_{T_r}$ 
is the 
mean square distance traveled on 
the quenched lattice during the time $T_r$, the diffusion coefficient on 
the stirred lattice will be 
 $D=\left<r^2\right>_{T_r}/2dT_r$. The case of local reorganization,
which is studied in this article, is 
more realistic, because the evolution of the network is continuous. The
 state of a bond evolves through a Poissonian process 
with a characteristic time $\tau$. 
At each iteration a conducting bond is cut with a 
probability  $1/(p\tau)$, and a randomly chosen non conducting bond becomes 
conducting, to insure that the proportion $p$ of conducting bonds is conserved.
No exact result is available for the dependence of the
diffusion coefficient $D$ on $p$ and $\tau$, except in some
particular one dimensional situations \cite{Garcia:1989}. Approximative
solutions of the problem in any dimension can be  obtained by means of 
a time-dependent version of the effective-medium approximation
developed in \cite{Harrison:1985}.

Here, we study the scaling of the diffusion coefficient $D$
in the vicinity of the percolation threshold $p_c$ of the quenched
network. Several different scaling formulas for $D(p-p_c,\tau)$ 
were proposed in the literature.
They were derived for models with slightly different
local evolution rules, but the details in the local
rules are not relevant for the critical behavior around $p_c$
\cite{Bug:1987}. As discussed below, our simulation
 results do {\em not} support current  predictions. We derive a new 
scaling formula for the diffusion coefficient, which we confirm
by extensive numerical simulations. The behavior at the percolation
 threshold is studied first, before we treat the general case of the behavior
around $p_c$. 

\begin{figure}
\centering
\epsfxsize=6.75cm
\leavevmode
\epsffile{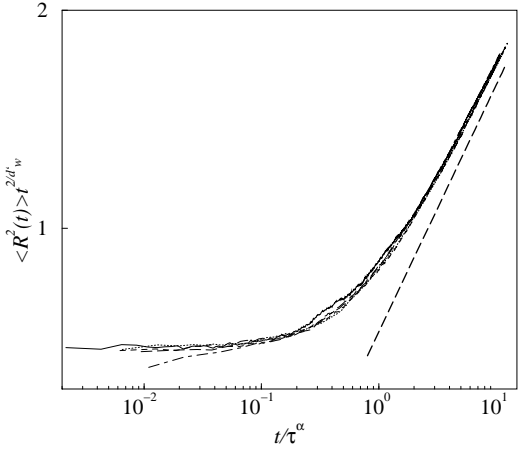}
\vskip1mm
\caption{Scaling function $\chi$ (Eq. 
(\ref{eq:scaling_seuil})) in two dimensions. Results obtained for 
$\tau=1.62 \times 10^6$ (solid line), $\tau=4.05 \times 10^{5}$ 
(dotted line), $\tau=1.8 \times 10^{5}$ (dashed line), $\tau=4.5 
\times 10^{4}$ (long dashed line), $\tau=5.0 \times 10^{3}$ (dot dashed line) 
with the blind ant algorithm. Asymptotic behavior $y\sim x^{1-2/d_w^{\prime}}$
 (bold long dashed line).}
\label{fig:scaling2D}
\end{figure}

\begin{figure}
\centering
\epsfxsize=6.75cm
\leavevmode
\epsffile{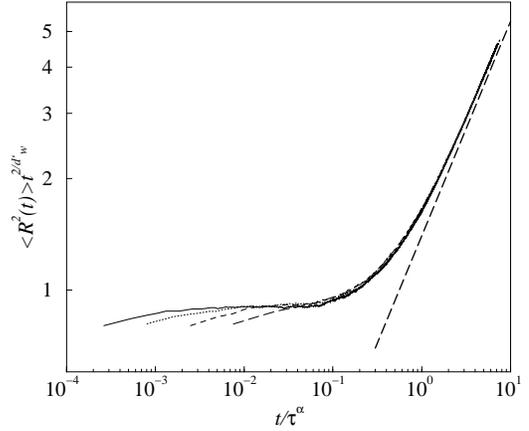}
\vskip1mm
\caption{Scaling function $\chi$ (Eq. 
(\ref{eq:scaling_seuil})) in three dimensions. Results obtained for 
$\tau=5.12 \times 10^{5}$ (solid line), $\tau=1.28 \times 10^{5}$ 
(dotted line), $\tau=3.12 \times 10^{4}$ (dashed line), $\tau=7.8 
\times 10^{3}$ (long dashed line) with the myopic ant algorithm. 
Asymptotic behavior $y\sim x^{1-2/d_w^{\prime}}$ (bold long dashed line)}
\label{fig:scaling3D}
\end{figure}

The mean square displacement in the vicinity of $p_c$ on a quenched 
percolation network is given by \cite{Havlin:1987}
\beq
\left<R^2\right> = 
t^{2/d_w^{\prime}}f\left[(p-p_c)t^{1/(2\nu+\mu-\beta)}\right],
\label{eq:scaling_static}
\eeq
where $d_w^{\prime}$ is the anomalous-diffusion exponent, 
$d_w^{\prime}=(2\nu+\mu-\beta)/(\nu-\beta/2)$ and
\[f(x)\sim \cases{x^{\mu}&as $x \to  \infty$ \cr 
(-x)^{-2\nu+\beta}&as $x\to -\infty$\cr const. &as $x \to 0$ . }\]
At early times, anomalous diffusion is observed. The crossover 
to a normal diffusion (if $p>p_c$) or to a localization regime
(if $p<p_c$) appears at a time of the order of $t_c
 \sim |p-p_c|^{\beta-2\nu-\mu}$, which is the only relevant timescale
of the problem. In the case of dynamically disordered lattices,
another timescale, related to the cluster rearrangement process, has to be
taken into account. We define this typical ``rearrangement'' time $t_r$ as
crossover time from anomalous to normal diffusion at the percolation 
threshold. It is only a function of the evolution rate $\tau$, and we assume a
 dependence in the form 
$t_r \sim \tau^{\alpha}$. 
The mean square displacement in the presence of dynamical
disorder is thus described by a scaling formula depending on two parameters,
$t/t_c$ and $t/t_r$ \cite{Kutner:1983}:
\beq 
\left<R^2\right> = t^{2/d_w^{\prime}}g\left[(p-p_c)t^{1/(2\nu+\mu-\beta)} ; 
t/\tau^{\alpha}\right].
\label{eq:scaling_dynamic}
\eeq
At the percolation threshold, $t_c$ diverges, and the
preceding expression reads
\beq
\left<R^2\right> = t^{2/d_w^{\prime}}\chi(t/\tau^{\alpha})\, ,
\label{eq:scaling_seuil}
\eeq
where $\chi(y)\sim const.$ as $y \to 0$  and $\chi(y)= Dy^{1-2/d_w^{\prime}}$
as $y\to \infty$. The diffusion coefficient $D$ is obtained in the limit
$t \to \infty$ by
\beq
D \sim \tau^{-\alpha\mu/(2\nu+\mu-\beta)}.
\label{eq:alpha}
\eeq
Eq. (\ref{eq:alpha}) contains an unknown parameter $\alpha$.
Several values of $\alpha$ were proposed in the literature.
In \cite{Bug:1987}, the problem was mapped on the 
continuous random walk, and the lower and upper bounds for $\alpha$ 
were predicted.
In \cite{Kutner:1983}  $\alpha = 1$ is considered. The only justification
for this value is the assumption that the global and local rearrangement
models have the same behavior. 
We have performed Monte Carlo simulations to evaluate $\alpha$ 
numerically. The diffusion coefficient can only be measured
for small values of $\tau$, where the crossover time $t_r$ is small.
In order to explore a broader range
of values, we have determined  $\alpha$  from the
finite size scaling relation (\ref{eq:scaling_seuil}). We measured 
$\left<R^2\right>$
for  $\tau$ between
 $5 \times 10^{3}$ and  $1.62 \times 10^{6}$ in two dimensions, and between
$7.8 \times 10^{3}$ and $5.12 \times 10^{5}$ in three dimensions. 
 In two dimensions the best data collapse with parameter $\alpha$ 
is obtained for 
$\alpha = 0.80\pm 0.02$ (Fig. \ref{fig:scaling2D}),
in three dimensions for $\alpha=0.79\pm 0.03$ 
(Fig. \ref{fig:scaling3D}). Identical results were obtained  with both
the myopic and the blind ant algorithms.

As a matter of fact, the value of $\alpha$ can be evaluated as a function of 
known critical
exponents using simple assumptions about the geometry of clusters.
Clusters are composed of well connected blobs, 
interconnected by singly connected  bonds (``red bonds'') 
\cite{Coniglio:1982}.  If a red bond is cut,
the cluster breaks into two parts. We argue that the crossover time
corresponds to a removal (or addition) of a red bond in the region visited
by the tracer particle. The red bonds are the only possible paths where
a particle can escape from a blob, hence they control the
 diffusion. 
For $t<t_r$ a particle visits on average a hypersphere
of a diameter $R \sim t^{1/d_w^{\prime}}$. The ``network'' of red bonds
is fractal, and their number inside the hypersphere grows as
$N_{\rm rb} \sim R^{1/\nu}$ \cite{Coniglio:1981}.
The crossover corresponds to the average time for the first 
of $N_{\rm rb}$ red bonds to be cut. Hence
 $t_r \sim \tau/N_{\rm rb} \sim \tau^{d_w^{\prime}/(d_w^{\prime}+1/\nu)}$, 
giving  
\beq
\alpha=\frac{d_w^{\prime}}{d_w^{\prime}+1/\nu}\ .
\label{alpha}
\eeq
In two dimensions, where $\nu=4/3$, $\beta=5/36$ \cite{Stauffer:1992} 
and $\mu=1.303$ \cite{Frank:1988} we obtain 
$\alpha = 0.802$. In three dimensions $\alpha=0.81\pm 0.06$
is obtained, using
$\nu=0.88\pm 0.02$, $\mu=2.003\pm 0.047$ \cite{Gingold:1990} 
and $\beta=0.405\pm 0.025$ \cite{Adler:1990}.
These values of $\alpha$
are in excellent agreement with numerical results. 
Relation (\ref{alpha}) predicts that $\alpha=1$ for $d \geq 6$, so
in this limit the local and the global reorganization rules lead to
the same scaling.

\begin{figure}
\centering
\epsfxsize=6.75cm
\leavevmode
\epsffile{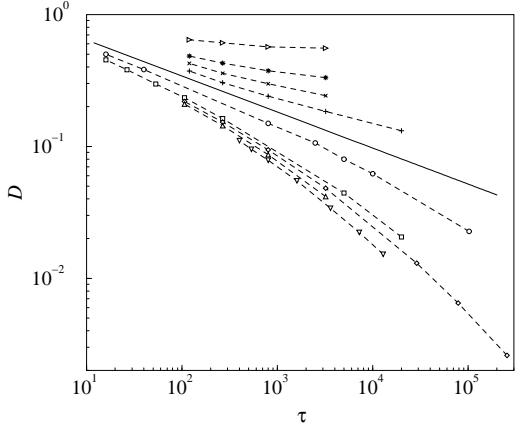}
\vskip1mm
\caption{Calculation of $D$ for different values of 
$|p-p_c|$ and
$\tau$. Results for $p=0.47$ $(\circ)$, $p=0.43$ $(\Box)$, 
$p=0.42$ $(\diamond)$, $p=0.41$ $(\triangle)$, $p=0.40$ $(\bigtriangledown)$,
  $p=0.53$ $(+)$, $p=0.56$ $(\times)$, $p=0.60$ $(*)$, $p=0.70$ 
$(\triangleright)$. Function (\ref{eq:alpha}) (solid line)}
\label{fig:D}
\end{figure}

Knowing the value of $\alpha$, the complete scaling law for $D$
in the vicinity of the percolation threshold can be deduced from
(\ref{eq:scaling_dynamic}). 
The ratio $t_c/t_r$ separates two different regimes. In the fast rearrangement
regime  ($t_c/t_r \gg 1$) a tracer particle does not see the finiteness of 
cluster sizes, hence the scaling of $D$ is given by (\ref{eq:alpha}).
In the slow rearrangement regime ($t_c/t_r \ll 1$) two cases have to
be considered. For $p>p_c$ and $\tau \to \infty$, known
results for the diffusion on the quenched network \cite{Havlin:1987} 
should be recovered, hence
\beq
D \sim |p-p_c|^{\mu}.
\label{eq:pg}
\eeq
For $p<p_c$ the situation is more complicated.
 At $t\approx t_c \ll t_r$, the network is not yet reorganized and
anomalous diffusion crossovers to a localization regime  on
a finite cluster exactly in the same way as for the quenched network.
The mean square displacement is thus 
$\left<R^2\right> \sim |p-p_c|^{\beta-2\nu}$.
For $t>t_c$ it grows as  
\[\left<R^2\right> \sim |p-p_c|^{\beta-2\nu}g^{\prime}\left[|p-p_c|
t^{1/(2\nu+\mu-\beta)} ; t/\tau^{\alpha}\right].\]
 For $t \to \infty$ a diffusive regime
is reached, and it is evident that $D \sim 1/\tau$ in this case.
Thus the scaling function $g^{\prime}$ behaves as 
$g^{\prime}[x,y] \sim x^{-a}y^{1/\alpha}$ for $x,y \to \infty$,
where the coefficient $a$ reads
\beq
a=(1/\alpha-1)(2\nu+\mu-\beta)=1-{\beta \over 2\nu}.
\label{eq:a}
\eeq
The final expression was obtained replacing $\alpha$ by (\ref{alpha}).
Then $a=0.948$ in two dimensions and $a=0.77\pm0.017$ 
in three dimensions.  
The scaling relation for $t>t_c$ can thus be written as a function of a unique
parameter
\beq
\left<R^2\right> \sim |p-p_c|^{\beta-2\nu}g^{\prime\prime}\left[\frac{t}
{|p-p_c|^{a}\tau}\right]\, ,
\label{eq:sr}
\eeq
with $f(y)\sim const$ for $y \to 0$, $f(y)\sim y$ for $y \to
\infty$. It is readily seen that the crossover time $t_c^{\prime}\equiv 
|p-p_c|^a \tau$ has itself a critical behavior near $p_c$ with an 
exponent $a$. 
This fact has been already predicted in \cite{Kerstein:1986} but a different
exponent $a=1$ was proposed. The scaling of $D$
in the slow rearrangement regime is simply deduced from (\ref{eq:sr}), 
\beq
D \sim \frac{|p-p_c|^{\beta-2\nu-a}}{\tau}. 
\label{eq:ajeto}
\eeq
The complete scaling law for $D$ consistent with 
(\ref{eq:alpha}) (\ref{eq:pg}) and (\ref{eq:ajeto}) reads
\beq
D=\frac{|p-p_c|^{\beta-2\nu-a}}{\tau}\phi\left[(p-p_c)
\tau^{1/(2\nu+\mu+a-\beta)}\right],
\label{eq:compl}
\eeq
with
\[\phi(z)\sim \cases{const.&as $z \to -\infty$ \cr |z|^{2\nu+a-\beta} &as 
$z\rightarrow  0$\cr z^{2\nu+\mu+a-\beta} &as $z \to \infty$ .\cr}\]

\begin{figure}
\centering
\epsfxsize=6.75cm
\leavevmode
\epsffile{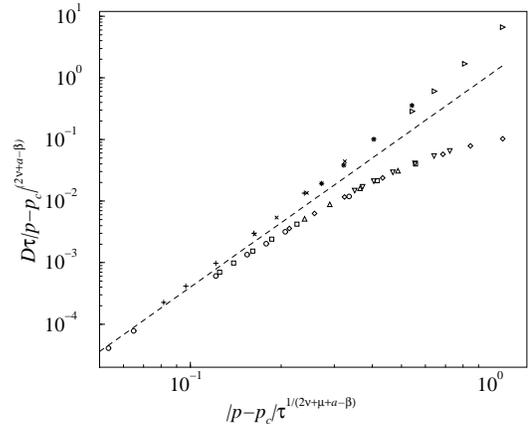}
\vskip1mm
\caption{Results of calculation of $D$ 
for different values of $|p-p_c|$ and
$\tau$ rescaled by (\ref{eq:compl}). Results for $p=0.47$ $(\circ)$, 
$p=0.43$ $(\Box)$, $p=0.42$ $(\diamond)$, $p=0.41$ $(\triangle)$, 
$p=0.40$ $(\bigtriangledown)$, $p=0.53$ $(+)$, $p=0.56$ $(\times)$, 
$p=0.60$ $(*)$, $p=0.70$ $(\triangleright)$. Function 
$f(z)\sim z^{2\nu+a-\beta}$ (dashed line)}
\label{fig:Dcol}
\end{figure}

To verify this relation, we have calculated the diffusion coefficient in 
two dimensions for different values of 
$\tau$ and for $p$ in the range $[0.4;0.47]$ and $[0.53;0.7]$, using the
algorithm of the myopic ant. Results are presented in Figure \ref{fig:D}.
They are well rescaled by the relation (\ref{eq:compl}) 
(Fig. \ref{fig:Dcol}). The best collapse seems to be 
reached for a slightly smaller value of $a$ ($a=0.9$) than predicted by 
(\ref{eq:a}) ($a=0.948$). However,  the collapse is not very sensitive 
on the precise value of $a$, because the slow rearrangement regime
is not explored in our range of ($\tau,p$). It is difficult to
attain this regime using a simple random walk, since 
the crossover time  to the diffusive behavior becomes too important 
for large values of $\tau$.
This is the reason why we used the following  algorithm to verify Eq. 
(\ref{eq:ajeto}).

\begin{figure}
\centering
\epsfxsize=6.75cm
\leavevmode
\epsffile{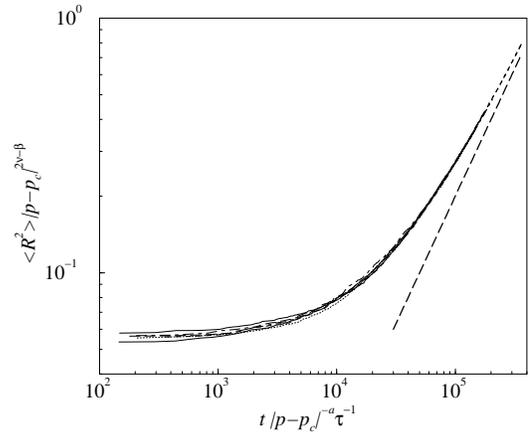}
\vskip1mm
\caption{Scaling function $g^{\prime\prime}$ 
(\ref{eq:sr}) in two dimensions. The data for $p=0.4$ (solid line), 
$p=0.41$ (dotted line), $p=0.42$ (dashed line), $p=0.43$ (long dashed line), 
$p=0.44$ (dot dashed line) and $p=0.46$ (solid line) Asymptotic behavior 
$y\sim x$ (bold long dashed line)}
\label{fig:slow2D}
\end{figure}

We start from a given site belonging to a cluster of $s$ sites. We suppose that
the evolution of the network is quasistatic : before the network is 
rearranged, the particle is thermalized, so that the
probability to find it on a given cluster site equals
$1/s$. Thus we assign at first the probability $1/s$ to each cluster site.
We then exchange one conducting bond with an 
insulating bond, find a new cluster distribution  and thermalize 
the probability distribution
on each cluster. We iterate this procedure
and measure the mean square displacement. The Hoshen-Kopelman
algorithm \cite{Hoshen:1976} was used to obtain the 
distribution of clusters. To get good 
statistics, an average over more than 2000 realizations was performed,
so we were limited to  networks of relatively small size  (up to
$400\times 400$ sites). Since the  diffusive regime is not attained on such
a small network, we used the finite size scaling 
formula (\ref{eq:sr}). We measured $\left<R^2\right>$ for $p$ ranging from
0.4 to 0.46. For higher values of $p$, clusters are too large, and much
bigger networks have to be used. The  data collapse
is obtained for $a=0.87\pm 0.05$ (Figure \ref{fig:slow2D}), that is
for a value slightly smaller than predicted by (\ref{eq:a}).
The same effect as in the case of the data collapse of $D(p-p_c,\tau)$ is
thus encountered.
The discrepancy is due to fact that we
are already out of the critical region, so corrections to the
exponents $\alpha$ and $d_w$ should be taken into account. For values
of $p$ far from $p_c$, the probability of having a large cluster,
corresponding to a long jump, grows more slowly than near $p_c$, and
 the growth of the diffusion coefficient with $p$ is thus also slower.

In conclusion, we have derived a new scaling law for the diffusion coefficient
in the case of a simple model of stirred percolation. The dependence of the
scaling exponents on the basic exponents of the percolation theory was found. 
We showed that the distribution of red bonds controls the transport in the
network. Results are supported by extensive numerical simulations. 
In the slow rearrangement regime for $p<p_c$ the diffusion coefficient scales
as $D \sim |p-p_c|^{s^{\prime}}$, where $s^{\prime}\doteq -2.1$ in three
dimensions. The value of the scaling exponent in microemulsions
($s^{\prime}\doteq 1.2$) 
\cite{Lagues:1979,Bhattacharya:1985,Cametti:1990}
thus cannot be explained by this simple model, as suggested earlier 
\cite{Grest:1986}. 
It is plausible that the difference
is due to interparticle interactions present in microemulsions.
They play an important role in the 
formation of clusters \cite{Safran:1985} and they might also
influence the dynamics of the reorganization of the environment.

I would like to thank J\'er\^ome Chave for many fruitful conversations
and Hugues Chat\'e for a careful reading of the manuscript. I also thank
Roger Bidaux and Marc A. Dubois for useful discussions.

\end{document}